\documentclass[prb,twocolumn,superscriptaddress]{revtex4-1}
\usepackage{graphicx}
\usepackage{dcolumn}
\usepackage{bm}
\usepackage{hyperref}
\usepackage[mathlines]{lineno}
\usepackage{textcomp}
\usepackage{natbib}
\usepackage{float}
\usepackage{amsmath}
\usepackage{amssymb}

\begin{document}

\title{Superconducting quantum point contact with split gates in the two dimensional LaAlO$_3$/SrTiO$_3$ superfluid}

\author{Holger Thierschmann}
\thanks{these authors contributed equally; h.r.thierschmann@tudelft.nl}

\author{Emre Mulazimoglu}
\thanks{these authors contributed equally; h.r.thierschmann@tudelft.nl}

\affiliation{Kavli Institute of NanoScience, Faculty of Applied Sciences, Delft University of
	Technology, Lorentzweg 1, 2628 CJ Delft, The Netherlands}

\author{Nicola Manca}
\affiliation{Kavli Institute of NanoScience, Faculty of Applied Sciences, Delft University of
	Technology, Lorentzweg 1, 2628 CJ Delft, The Netherlands}

\author{Srijit Goswami}
\affiliation{Kavli Institute of NanoScience, Faculty of Applied Sciences, Delft University of
	Technology, Lorentzweg 1, 2628 CJ Delft, The Netherlands}
\affiliation{QuTech, Delft University of Technology, Lorentzweg 1, 2628 CJ, Delft, The Netherlands}

\author{Teun M. Klapwijk}
\affiliation{Kavli Institute of NanoScience, Faculty of Applied Sciences, Delft University of
	Technology, Lorentzweg 1, 2628 CJ Delft, The Netherlands}
\affiliation{Physics Department, Moscow State Pedagogical University, Moscow 119991, Russia.}

\author{Andrea D. Caviglia}
\email{a.d.caviglia@tudelft.nl}
\affiliation{Kavli Institute of NanoScience, Faculty of Applied Sciences, Delft University of
	Technology, Lorentzweg 1, 2628 CJ Delft, The Netherlands}

\date{\today}
\begin{abstract}
One of the hallmark experiments of quantum transport is the observation of the quantized resistance in a point contact formed with split gates in GaAs/AlGaAs heterostructures \cite{van1991quantum,wharam1988one}. Being carried out on a single material, they represent in an ideal manner equilibrium reservoirs which are connected only through a few electron mode channel with certain transmission coefficients \cite{van1991quantum}. It has been a long standing goal to achieve similar experimental conditions also in superconductors \cite{beenakker1991josephson}, only reached in atomic scale mechanically tunable break junctions of conventional superconducting metals, but here the Fermi wavelength is so short that it leads to a mixing of quantum transport with atomic orbital physics \cite{scheer1997conduction}. 
	Here we demonstrate for the first time the formation of a superconducting quantum point contact (SQPC) with split gate technology in a superconductor, utilizing the unique gate tunability of the two dimensional superfluid at the LaAlO$_3$/SrTiO$_3$ (LAO/STO) interface \cite{reyren2007superconducting,caviglia2008electric,goswami2016quantum}.
	When the constriction is tuned through the action of metallic split gates we identify three regimes of transport: (i) SQPC for which the supercurrent is carried only by a few quantum transport channels. (ii) Superconducting island strongly coupled to the equilibrium reservoirs. (iii) Charge island with a discrete spectrum weakly coupled to the reservoirs.
	Our experiments demonstrate the feasibility of a new generation of mesoscopic all-superconductor quantum transport devices.
\end{abstract}

\maketitle

\begin{figure}
	\centering
	\includegraphics[width=1\linewidth]{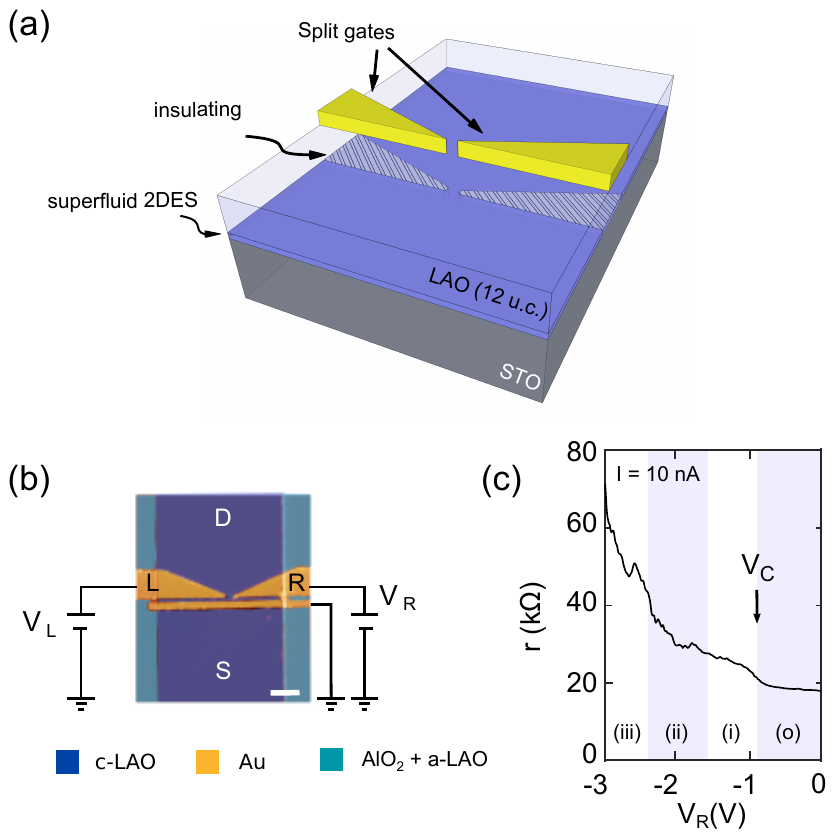}
	\caption{\textbf{Split gate device: (a)} Generalized sketch of the device. At the interface between the STO substrate and the the 12 unit cell (u.c.) layer of crystalline LAO (c-LAO) the superconducting 2DES (blue) is formed, which can be tuned insulating (shaded blue) locally under the split gates (yellow), thus forming a superconducting constriction. \textbf{(b)} False color AFM image showing the device layout. The potential of the split gates (yellow) L and R is controlled with the voltages V$_L$ and V$_R$, respectively. V$_L$ is kept at -1V. The conductive 2DES is formed in regions with c-LAO (blue). In areas which are protected with an AlO$_2$ hard mask LAO growth is amorphous (a-LAO, turquoise). The thin gate spanning the channel is not used in the experiments. The scale bar corresponds to 1 $\mu$m. \textbf{(c)} High bias (I=10 nA) differential resistance $r$ as a function of V$_R$. (o) - (iii) indicate the different regimes of transport (see text). V$_c$ denotes the formation of the constriction.
	}
	\label{fig:1}
\end{figure}

Various attempts have been made to combine the desired gate-tunability of the low electron density semiconductor with the use of conventional superconductors. However, these hybrid devices have introduced, compared to the GaAs/AlGaAs normal quantum transport case, the very important and yet very difficult to control influence of the interface between the two dissimilar materials \cite{takayanagi1995observation}. This makes the results dependent on the complexities of the proximity effect and thus complicates their interpretation.
In principle, a new path has become available when it was discovered that in the two dimensional electronic system (2DES) at the LAO/STO interface superconductivity becomes suppressed when the electron density $n$ is reduced below a critical value $n_\text{c}$, for example by means of a gate voltage \cite{caviglia2008electric,goswami2015nanoscale}.  
The Fermi-wavelength $\lambda_F$ in this system can be as large as $ 30 \text{ to } 50 $ nm \cite{tomczyk2016micrometer} and ballistic transport in the normal state has been demonstrated \cite{gallagher2014gate,tomczyk2016micrometer}. The superconducting coherence length is about $\xi = 100 $ nm \cite{reyren2007superconducting}. This corresponds to spatial dimensions which are commonly achieved with present day lithography techniques. The creation of an SQPC with split gates in LAO/STO should therefore be within reach [cf. Fig.\ref{fig:1}(a)]. 
This approach can also offer insight into the nature of superconducting pairing at oxide interfaces.  
Unconventional pairing was recently suggested \cite{stornaiuolo2017signatures,fidkowski2013magnetic} in light of a number of experimental observations, including strong spin orbit coupling \cite{caviglia2010tunable}, co-existence of ferromagnetism and superconductivity \cite{bert2011direct,dikin2011coexistence,li2011coexistence}, indications for electron pairing without macroscopic phase coherence \cite{cheng2015electron,cheng2016tunable,richter2013interface,tomczyk2016micrometer} and a non-trivial relation between the critical temperature $T_\text{c}$ and charge carrier density $n$ \cite{caviglia2008electric,richter2013interface}.

Here we present experiments that demonstrate the formation of an SQPC with split gates in the LAO/STO superfluid.
Our sample is fabricated following the procedure described by Goswami et al. ~\cite{goswami2016quantum} [see Methods].
Figure~\ref{fig:1}(b) presents a false color atomic force microscope (AFM) image of the device layout. The metallic split gates (yellow) L and R cover the full width of the 5 $\mu$m wide 2DES (blue), except for a 150 nm region at its center.
Transport experiments are performed in a current bias configuration (unless stated otherwise) at temperature $T_\text{base} <$ 40 mK. The resisitvly measured transition to the superconducting state is observed at $T_c \approx $100 mK [see Methods and SI].
Because of gate history effects, we carry out the experiment by putting electrode L on a fixed gate voltage ($V_\text{L}$ = -1V) to ensure depletion and we tune the constriction by only varying the voltage $V_R$ applied to gate R [see SI].

We expect the following scenario: When $V_\text{R}$ is changed towards negative values the charge carrier density $n$ gets reduced locally underneath the gate and gets closer to the critical density $n_c$ at which superconductivity becomes suppressed. At a certain gate voltage $V_\text{R} = V_\text{c}$ the condition $n = n_\text{c}$ is reached and a supercurrent can flow only through the constriction between the tips of gates, thus forming a weak link between the superconducting reservoirs. Outside this weak link, under the gates, the system acts as an insulator \cite{caviglia2008electric}. The number of transport modes available in the weak link is determined by its effective width.
The constriction width is reduced when V$_R$ is further decreased and therefore the number of transmission channels decreases which is expected to lead to a step-wise reduction of the critical current $I_\text{c}$ \cite{beenakker1991josephson}. For V$_R \ll V_c$ transport will be dominated by a low transmissivity and the current is pinched off.

\begin{figure}
	\centering
	\includegraphics[width=1\linewidth]{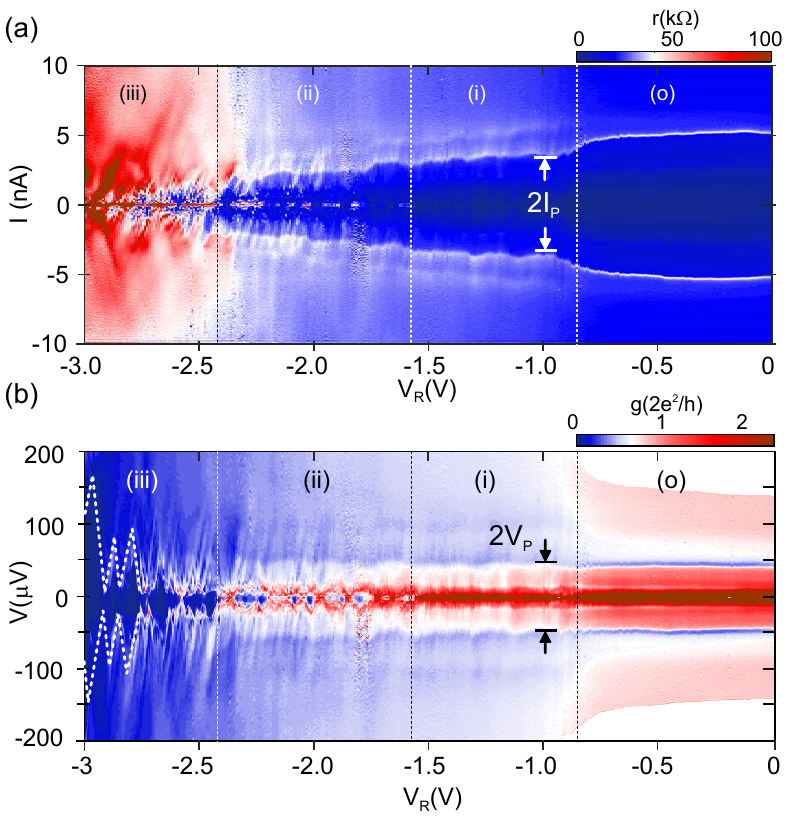}
	\caption{ \textbf{Transport regimes in the constriction} \textbf{(a)} Differential resistance $r$ versus current $I$ for gate voltage $V_\text{R} = 0$ to $-3$V. Gate L is kept at $V_L = -1$ V. The four regimes of transport (o) to (iii) are indicated. I$_P$, reminiscent of the critical current $I_c$ of the weak link, is indicated. \textbf{(b)} Same data as in (a) but with voltage drop $V$ on the vertical axis and differential conductance $g$ represented by the color scale. V$_P$ denotes the voltage drop at I$_P$. Dotted lines in regime (iii) indicate conductance diamonds.}
	\label{fig:overview}
\end{figure}

In order to study this scenario, we record a series of $V$-$I$ curves and vary $V_\text{R}$ from 0 to -3V.
Panoramic overviews of the results are given in Fig.~\ref{fig:overview} (a) and (b) in color plots. Figure \ref{fig:overview}(a) presents the differential resistance $r=$d$V/$d$I$ and Fig.~\ref{fig:overview}(b) shows the differential conductance $g=$d$I$/d$V$ with the current $I$ and voltage drop $V$ on the vertical axis, respectively.
It can be seen that the constriction undergoes four different regimes of transport [labelled (o) and (i) to (iii) in the figures] as $V_\text{R}$ is varied from 0 to -3 V.
Regime (o) 
(ranging from $V_\text{R}$ = 0 to -0.9 V) 
corresponds to the open current path configuration with $V_\text{R}>V_\text{c}$. 
A sharp peak in $r$ is visible at $I = \pm$ 5 nA, labelled $I_P$, which is reminiscent of a critical current $I_c$. Correspondingly, a dip occurs in $g$ at $V_\text{p}=\pm 44~\mu$V (Fig.\ref{fig:overview}(b)).
At V$_R = V_\text{c} \approx -0.9$ V the critical density $n_\text{c}$ is reached. Here $I_p$ drops significantly because the current path becomes confined. At high currents in Fig.\ref{fig:overview}(a), as shown for $I = 10$ nA in Fig.\ref{fig:1}(c), this point of confinement is apparent in a step increase in $r$, similar to the well-known behavior in semiconductor heterostructures \cite{zheng1986gate}. It marks the transition to regime (i)
($V_\text{R}$ = -0.9 to -1.6 V).
In this regime $I_P$ decreases when $V_\text{R}$ is reduced indicating the gate tunable weak link. 
In regime (ii) ($V_\text{R}$ = -1.6 to -2.4 V) regions of high resistance at zero bias appear and disappear periodically. As we will show below, this can be attributed to the emergence of a conductive island which dominates transport through the constriction. In regime (iii) ($V_\text{R}$ = -2.4 to -3 V) the device always exhibits a high resistance at zero bias. Figure \ref{fig:overview}(b) reveals that this regime is controlled by conductance diamonds (indicated with dashed lines).

\begin{figure*}
	\centering
	\includegraphics[width=0.8\linewidth]{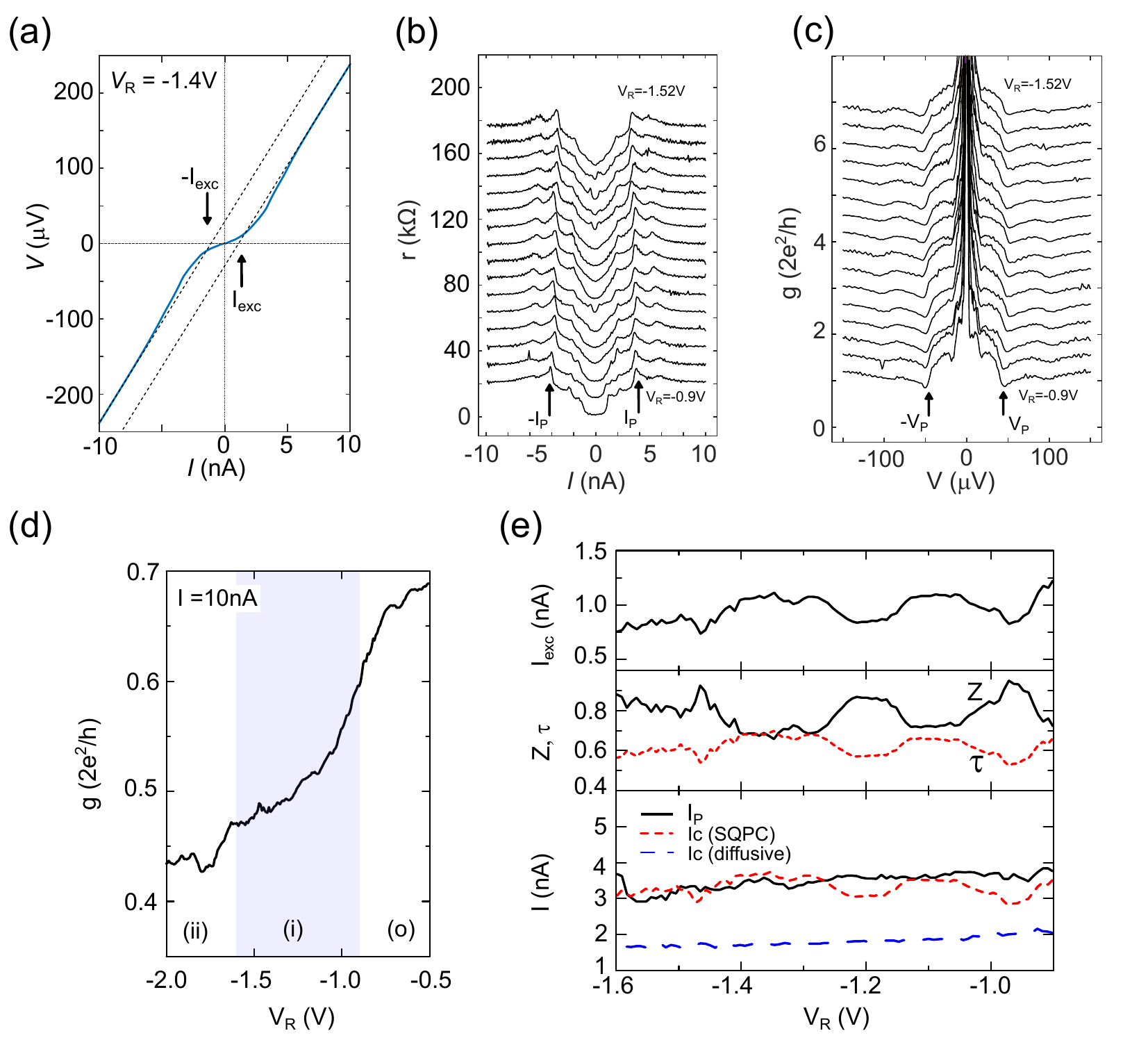}
	\caption{\textbf{Regime (i): SQPC.} \textbf{(a)} representative V-I curve in regime (i) ($V_R = -1.4$ V). The rounded supercurrent is clearly visible. The excess current is denoted I$_\text{exc}$. \textbf{(b)} $r$ versus $I$ for different V$_R$. I$_p$, inferred from the sharp peak in $r$, is indicated. The curves are offset by 10 k$\Omega$ \textbf{(c)} $g$ versus V for different V$_R$. V$_P \approx 44~\mu$V is indicated. The curves are offset by 0.2 (2$e^2/h$). \textbf{(d)} $g$ for I=10 nA. \textbf{(e)} top panel: $I_\text{exc}$ as a function of V$_R$. middle panel: Barrier strength Z (solid, black line) and normal state transmission $\tau$ (dashed, red line) as obtained from BTK. Bottom panel: $I_P$ (solid, black line) and critical current I$_c$ as expected for an SQPC (short, red dashes). For comparison, I$_c$ expected for a diffusive junction with conductance $g$ is shown (blue dashes).}
	\label{fig:SQPC}
\end{figure*}

Let us start the discussion with the weak link regime (i). Here we observe a rounded supercurrent and an excess current $I_\text{exc} \approx$ 1nA [Fig.\ref{fig:SQPC} (a)].  $I_P$, reminiscent of the critical current $I_c$, changes from 3.7 to 3.0 nA [Fig.\ref{fig:SQPC}(b)] when V$_\text{R}$ is varied.  The voltage $V_P = 44~\mu$V [cf. Fig.~\ref{fig:SQPC}(c)] can be related to the superconducting gap $V_P/2 \approx \Delta \approx 22 ~\mu$eV, which is compatible with the value inferred from the resistively measured $T_c$, $\Delta_{Tc}=1.76k_BT_c = 15~\mu$eV. The high bias conductance $g_\text{n}$ [Fig.~\ref{fig:SQPC}(d)] is of the order of half the quantum of conductance, changing with V$_R$ from 0.6 to 0.47 (2$e^2$/h) [20 to 28 k$\Omega$]. 
As shown by Monteiro et al. \cite{monteiro2017side} the phase correlation length in our 2DES is about 170 nm, whereas the lithographically determined channel-width is about 150 nm. It is therefore reasonable to interpret the data from a quantum transport perspective. For low carrier densities the Fermi wavelength $\lambda_F$ is several 10 nm \cite{tomczyk2016micrometer}. This and the relatively low value of $g_\text{n}$ suggest that we have only a few modes with a finite transmissivity in the channel. 
With increasing $V_R$ we do not observe the expected quantum transport step-like features in $g_\text{n}$, although the trace in Fig.\ref{fig:1}(e) is obviously not monotonous.
In order to extract the transmissivity of the weak link we calculate from $I_\text{exc}$ and $g_\text{n}$ the barrier strength Z as a function of V$_R$ using the BTK-formalism for an S-S interface \cite{blonder1982transition}. Z is related to the normal state transmission probability $\tau$ by $\tau = (1+Z^2)^{-1}$. In this manner we obtain $Z\approx 0.8$ and, correspondingly, $\tau \approx 0.6$ [Fig.\ref{fig:SQPC}(e)]. Comparison with the measured g$_\text{n}$ thus suggests a total mode conductance of $2e^2/h$, such that $g(\tau=0.6) = 0.6\times2e^2/h$, close to the measured values. If we follow recent experiments by Gallagher et al. \cite{gallagher2014gate} who observed e$^2$/h modes in a normal state QPC, we could also consider only one mode with a higher transmissivity. However, this would require a re-analysis of the excess current based on an unconventional order parameter.

If we continue the discussion in the conventional picture, for a SQPC with perfect transmission ($\tau =1$) Beenakker and van Houten \cite{beenakker1991josephson} found that the critical current is given by $I_c=N e\Delta(\hbar)^{-1}$, where $N$ was chosen to represent the number of spin degenerate modes (which contribute each 2$e^2/h$ to the normal conductance). Using this relation and including the obtained $\tau$ as a pre-factor, we can calculate the maximum supercurrent expected for our device, which yields $I_c \approx 3$~nA. This is in good agreement with the measured $I_P$, as can be seen in the bottom panel in Fig.~\ref{fig:SQPC}(e). For comparison we also plot the expected $I_c$ for a diffusive junction \cite{beenakker1992three}, which clearly gives much smaller values. 
The critical current $I_c\approx 3$ nA implies a Josephson coupling energy $E_J=  6.2~\mu $eV. This is comparable to the bath temperature, $k_BT_\text{base}=3.4~ \mu $eV. Therefore, as for the few-mode atomic scale point contacts \cite{goffman2000supercurrent,chauvin2007crossover}, the supercurrent is rounded.

\begin{figure}
	\centering
	\includegraphics[width=1\linewidth]{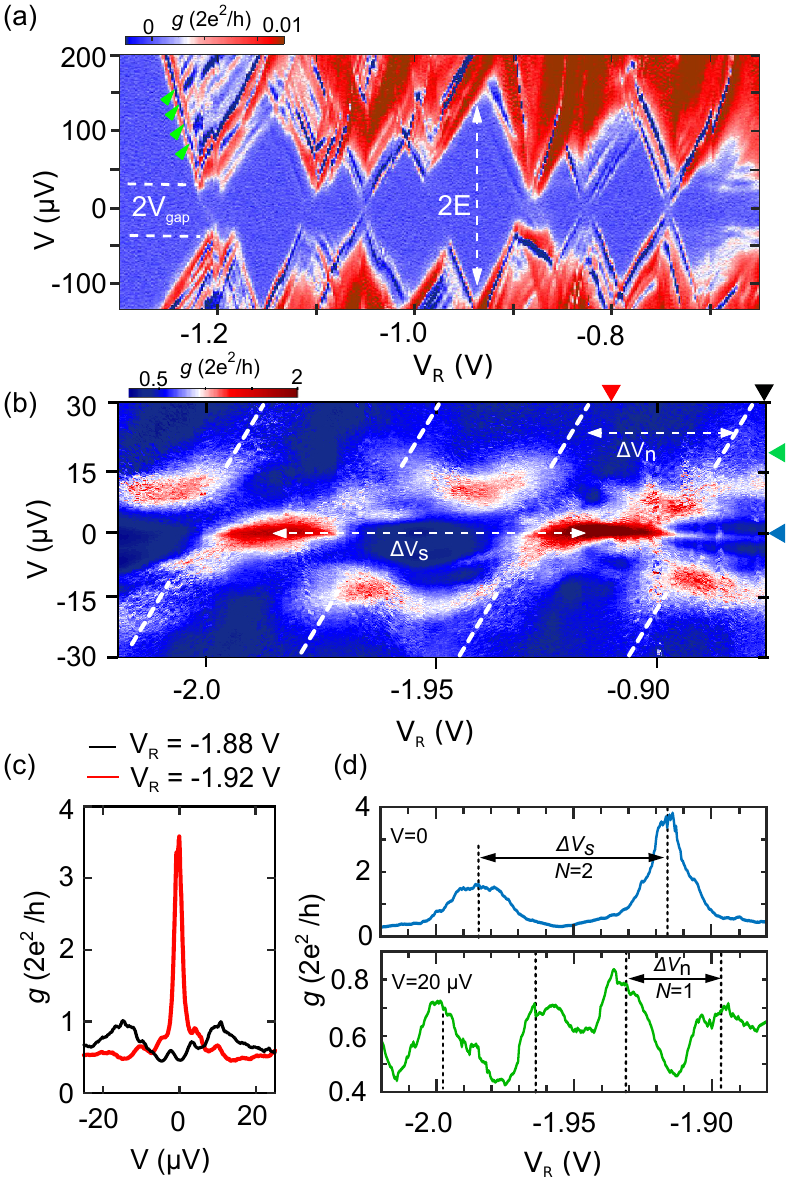}
	\caption{\textbf{Regimes (iii) and (ii): Conductive island.} \textbf{(a)} Conductance diamonds measured in regime (iii) with voltage bias V on the vertical axis. $E$ denotes the addition energy of the island. The diamonds exhibit a gap of $V_\text{gap} \approx \pm$ 30 $\mu V$. Green arrows indicate signatures of excited states due to quantum confinement. Note that due to gate history effects, this regime occurs at a lower V$_\text{R}$ range. \textbf{(b)} $g$ measured regime (ii). For small voltages $g$ exhibits peaks with periodicity $\Delta V_s = $ 70 mV. For $V>15~\mu $V the periodicity increases by a factor 2,  $\Delta V_n =$ 35 mV (dashed white lines). Black, red triangles and blue, green triangle indicate the line cuts shown in (c) and (d), respectively. \textbf{(c)} vertical line cuts from (b) at V$_\text{R}$ = -1.88 V and V$_\text{R}$ = -1.92 V. \textbf{(d)} Horizontal line cuts from (b) at V=0 (top panel) and V$= 20~ \mu$V (bottom panel). The change in periodicity by a factor 2 suggests a change in number of transferred charges from N=2 to N=1, indicative for a superconducting island. 
	}
	\label{fig:dot}
\end{figure}

Let us now turn to the regime of conductance diamonds (CDs), regime (iii). Figure \ref{fig:dot}(a) presents a detailed measurement of $g$ in this region. Note that this measurement was carried out in a voltage bias configuration.
We observe a series of CDs whose size $E$ on the (vertical) voltage axis is of the order of $80$ to $150~\mu$V. 
In gating experiments with non-superconducting materials, for instance in narrow semiconductor channels or graphene nano ribbons, CDs are known to occur in the low density limit because of puddles of charge carriers which form due to small inhomogeneities in the potential landscape, thus leading to quantum dot-like transport behavior \cite{staring1992coulomb,mceuen1999disorder,escott2010resonant,zwanenburg2013silicon,todd2008quantum,liu2009electrostatic}.
From this analogy we infer that in regime (iii) the superfluid inside the constriction is at the transition to full depletion.
The size of the CDs directly reflects the addition energy $E$ that has to be paid in order to change the island occupation number and thus, to enable transport. $E$ is composed of various contributions of which the most dominant ones typically are the Coulomb charging energy $U = Ne^2(2C_\Sigma)^{-1}$ [with $C_\Sigma$ being the total capacitance of the island and $N$ the number of charges to be added or removed] and the energy level quantization due to quantum confinement $\delta\varepsilon$.  
For quantum dots in LAO/STO \cite{maniv2016tunneling,cheng2015electron,cheng2016tunable}, Coulomb contributions are small because the STO substrate exhibits an extremely large dielectric constant $\epsilon_r=25000$ at low T and for small electric fields \cite{neville1972permittivity} which suppresses Coulomb repulsion. For our device, however, the fields originating from the split gates can not be neglected \cite{monteiro2017side}. We have performed simulations of the dielectric environment in the region surrounding the constriction using finite element techniques [see SI]. Our results indicate that the geometry of the gates leads to a strong field focusing effect which reduces $\epsilon_r$ in the constriction such that Coulomb repulsion becomes relevant. The numerical simulations yield charging energies of $U\approx 100~\mu$eV for an island with $\sim$ 50 nm radius, compatible with our experiment. The data in Fig.~\ref{fig:dot}(a) further show signatures of transport through excited states originating from quantum confinement, as can be seen from the fine structure of conductance lines parallel to the diamond edges between two adjacent diamonds [green arrows in Fig.\ref{fig:dot}(a)] \cite{hanson2007spins,zwanenburg2013silicon}. This allows us to estimate $\delta\varepsilon \approx 10 \text{ to } 20~ \mu$eV, which would lead to an island size of $\sim$ 80 nm, similar to the size obtained from the finite element simulations of the electrostatic properties. These values are also compatible with the electronic inhomogeneities typically observed in LAO/STO, which correlate with structural effects \cite{bert2011direct,honig2013local,kalabukhov2009cationic}.

The island couples to superconducting reservoirs, which can be inferred from the voltage gap $V_\text{gap} \approx \pm 30~ \mu$V that separates the CDs in positive and negative bias direction \cite{tuominen1992experimental,de2010hybrid}. As expected, $V_\text{gap}$ vanishes when a perpendicular magnetic field B=1T is applied [see SI]. 
We further observe pronounced negative differential conductance (NDC) along the edges of the CDs, which can be related to the sharp changes in density of states in the superconducting reservoirs around $\pm \Delta$. Since NDC occurs symmetrically for both positive and negative bias, we conclude that both reservoirs exhibit a superconducting energy gap [see SI].
When we compare the value of $V_\text{gap}=2\Delta$ with the superconducting gap in the reservoirs, $\Delta \approx 22~\mu$eV, we obtain reasonable agreement.
We note that in this regime (iii) the level spacing of quantum states on the island is of the same order as the superconducting gap, $\delta \varepsilon\sim\Delta$. We are therefore in the limit of Anderson's criterion of superconductivity at small scales ($\delta \varepsilon < \Delta$) \cite{anderson1959theory,van2001superconductivity}.

Finally we turn to the strong coupling regime (ii) [Fig.~\ref{fig:dot}(b)]. The pattern of gapped CDs is not visible here. Instead we observe zero bias conductance peaks which are of the order of the quantum of conductance, $g \geq (2e^2/h)$, [cf. Fig.\ref{fig:dot}(c), red curve]. They alternate with regions where $g$ is suppressed. This suggests that the island is more transparent in this regime, allowing for Cooper pair transport \cite{tinkham1996introduction} at zero bias. 
The peaks in $g$ occur periodically in $\Delta V_\text{R}$, with a periodicity $\Delta V_\text{s} = 70$ mV [Fig.~\ref{fig:dot}(d), top panel]. Above a certain bias voltage $V\approx \pm 15 \mu$V, the periodicity changes by a factor 2, $\Delta V_\text{n} = 35$ mV [Fig.~\ref{fig:dot}(d), bottom panel]. This suggests that the parity of the island influences its energy state, as expected for a superconducting island \cite{tuominen1992experimental}. In its ground state the island hosts Cooper pairs (even parity) and thus exhibits a charging energy $2U$, reflecting the Cooper pair's charge 2$e$ ($N$=2). Above a critical bias voltage the odd-parity state becomes available for quasi particles in the reservoirs thus enabling single electron transport across the island ($N$=1). This results in period doubling of the Coulomb blockade oscillations. Our data therefore suggest that in the strong coupling regime (ii) the island is in a superconducting state, thus forming a superconducting quantum dot (SQD).

We conclude that we have realized for the first time a superconducting quantum point contact with a split gate technique, of which the superconducting and normal transport is independent of unknown material interfaces. The present technology can serve as a basis for future experiments which will make it possible to evaluate the microscopic properties of the LAO/STO interface superconductivity and the properties of genuine superconducting quantum point contacts as originally envisioned \cite{beenakker1991josephson}. It may furthermore enable the investigation of nano scale superconductivity in few electron quantum dots.

\section*{ Acknowledgments }

We thank L.M.K. Vandersypen for comments on our manuscript.
This work was supported by The Netherlands Organisation for Scientific Research (NWO/OCW) as part of the Frontiers of Nanoscience program, the Dutch Foundation for Fundamental Research on Matter (FOM) and the European research council (METIQUM, grant no. 339306). T.M.K. further acknowledges support from the Ministry of Education and Science of the Russian Federation under Contract No. 14.B25.31.007.

\section*{Author contributions}
S.G. T.M.K. and A.D.C. conceived the experiment. E.M. fabricated the samples. E.M. and H.T. carried out the experiments. H.T. and T.M.K analyzed the data with input from E.M.  N.M. carried out the finite element simulations. H.T., E.M. and T.M.K  wrote the manuscript. All authors commented on the manuscript.
A.D.C. supervised the project.


\section{Methods}

\subsection{Device Fabrication}
We use single crystal TiO$_2 $ terminated, (001) oriented  SrTiO$_{3} $ (Crystec  {\textcopyright} GmBH) as a substrate without further modification. The fabrication involves three electron beam lithography steps (EBL). The first EBL defines the positions of reference markers which are obtained by Tungsten \textit{(W)} sputtering and consecutive lift-off. The second EBL step patterns the geometry of the device: Those regions which are to remain insulating are covered with 20 nm of sputtered AlO$_{2}$ (lift-off process in warm (50 \textcelsius) acetone). Next, the LaAlO$_{3}$ (LAO) layer is grown by means of pulsed laser deposition (PLD) at 770 \textcelsius~with an O$_{2}$ pressure of $p_\text{O2}= 6 \times 10^{-5}$ mbar. Only in those regions which are not covered by the AlO$_2$ hard mask growth is crystalline such that the STO surface is covered with a 12 unit cell (5 nm) LAO layer, giving rise to the 2DES at the interface. In all other regions the AlO$_2$ mask prevents the formation of the 2DES and the LAO layer is amorphous. Growth is monitored in-situ by reflection high energy electron diffraction  \textit{(RHEED)} which confirms layer-by-layer growth. After LAO deposition, the sample is annealed for one hour at 600 \textdegree C and at a pressure of $p_\text{O2}=300$ mbar in order to suppress the formation of O$_{2}$ vacancies. 
The final EBL step defines the pattern of gate electrodes. Polymer residuals are removed with an Oxygen plasma. Evaporation of 100 nm gold \textit{(Au)} is followed by gentle lift-off in acetone. The sample is mounted in a chip carrier with silver paint, serving as a back gate. Ultrasonic wedge bonding provides Ohmic contacts to the 2-dimensional electron system (2DES).  

\subsection{Electrical measurement setup and device characterization}
All measurements (unless stated otherwise) are performed using dc electronics, with the current sourced at reservoir S of the sample and drained at reservoir D. The resulting voltage drop V is probed at separate contacts in the respective reservoirs. The dilution refrigerator is equipped with copper powder filters, which are thermalized at the mixing chamber, and Pi-filters at room temperature. 

The carrier density in the 2DES is adjusted globally by applying a negative back gate voltage $V_\text{BG} =  -1.875$~V, which corresponds to a reduced density compared to $V_\text{BG}=0$. We determine the carrier density from Hall measurements performed at 300 mK using voltage probes on opposite sides of the reservoir with width w = 150~$\mu$m. The longitudinal resistance is determined from voltage measurements between probes separated by l= 112.5 $\mu$m. This yields a carrier density $n \approx 3 \times 10^{13}$ cm$^{-2}$ and a mobility $\mu \approx$ 800 cm$^2$(Vs)$^{-1}$. For this carrier density we observe the resistively measured superconducting transition at $T_c\approx 100$ mK, which corresponds to a BCS gap $\Delta_{Tc} = 15~\mu$eV.   
\newpage

\bibliography{Bibliography_full}

\newpage

\mbox{}
\thispagestyle{empty}
\newpage

\appendix\onecolumngrid

\section{Supplementary Information}

\subsection{Device Fabrication}

\begin{figure}[H]
	\centering
	\includegraphics[width=0.9\linewidth]{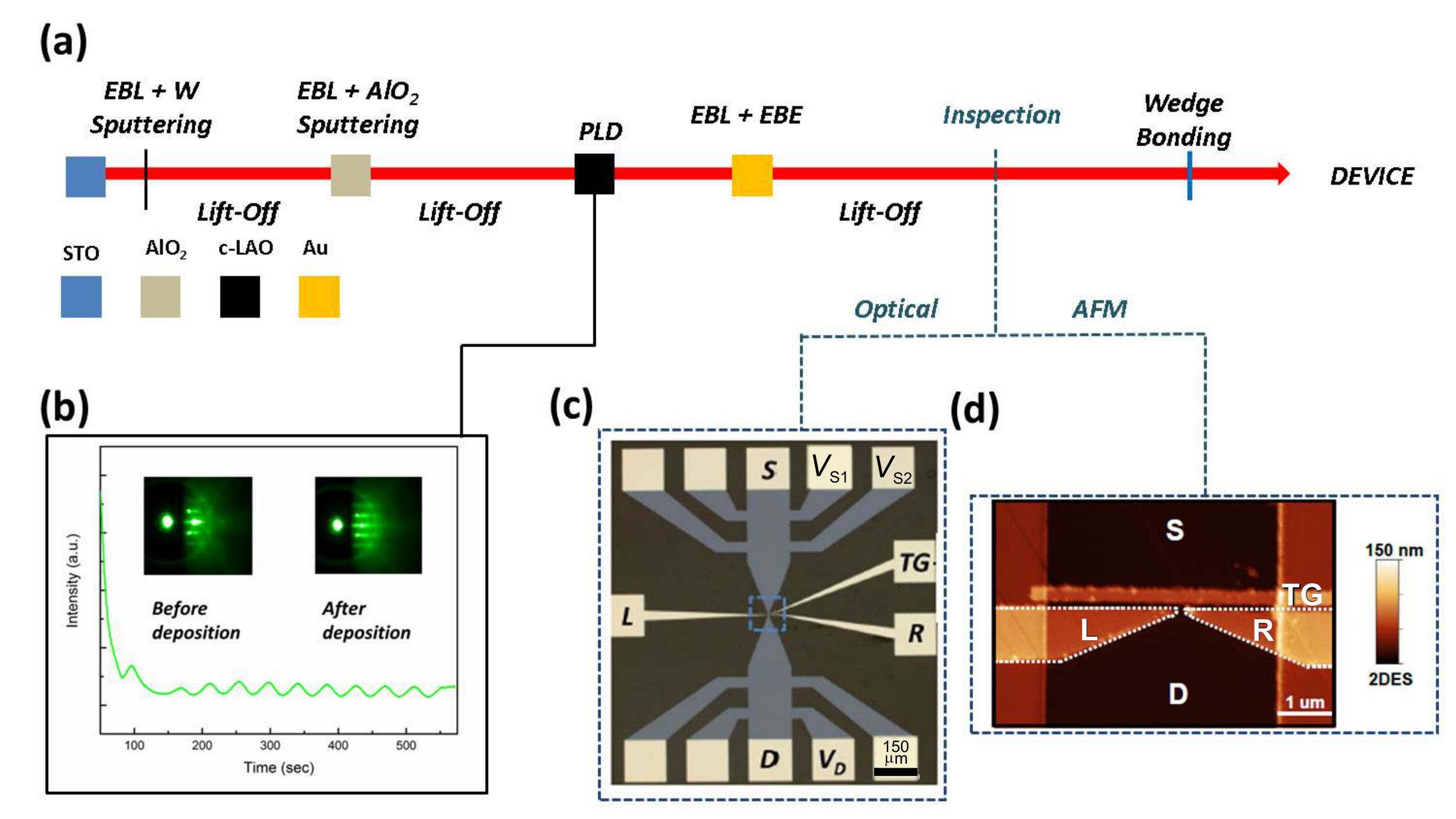}
	\caption{\textbf{Device fabrication:} \textbf{(a)} Fabrication flow. The color legend indicates the deposited materials at the different stages. \textbf{(b)} The in-situ RHEED oscillations monitored during growth. They confirm layer-by-layer growth. The insets show the RHEED diffraction pattern before and after the deposition of the LAO layers. \textbf{(c)} Optical image of the device after the lift off process. Ohmic contacts and the gate electrodes are labelled. \textbf{(d)} Atomic Force Microscopy (AFM) image showing the top gate architecture of the device.}
	\label{fig:SI_Fab}
\end{figure}

\newpage

\subsection{Resistively measured T$_c$}

\begin{figure}[H]
	\centering
	\includegraphics[width=\linewidth]{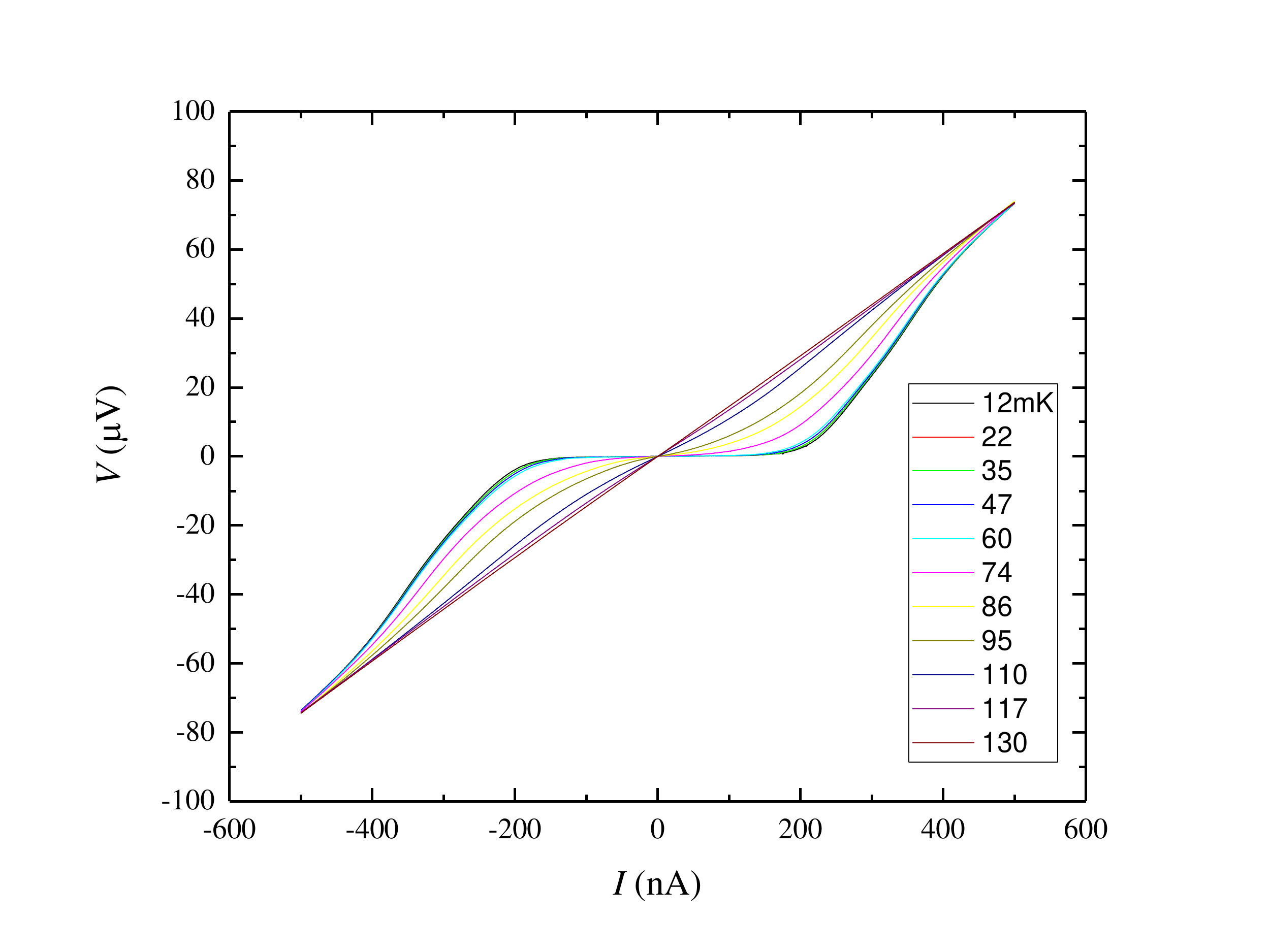}
	\caption{\textbf{Voltage-current curves for various temperatures,} measured in the reservoirs of the device using contacts S and D as source and drain for the current and V$_{S1}$ and V$_{S2}$ to probe the resulting voltage drop $V$. We obtain $T_C \approx 100 $ mK if we use $R_\text{TC}$ = $R(130$ mK)/2 to define $T_C$. Using the BCS equation for the superconducting gap $\Delta$, this yields $\Delta = 1.76k_BT_c \approx 15~\mu$eV}
\end{figure}

\newpage

\subsection{Additional data on the weak link regime (i)}

\begin{figure}[H]
	\centering
	\includegraphics[width=0.5\linewidth]{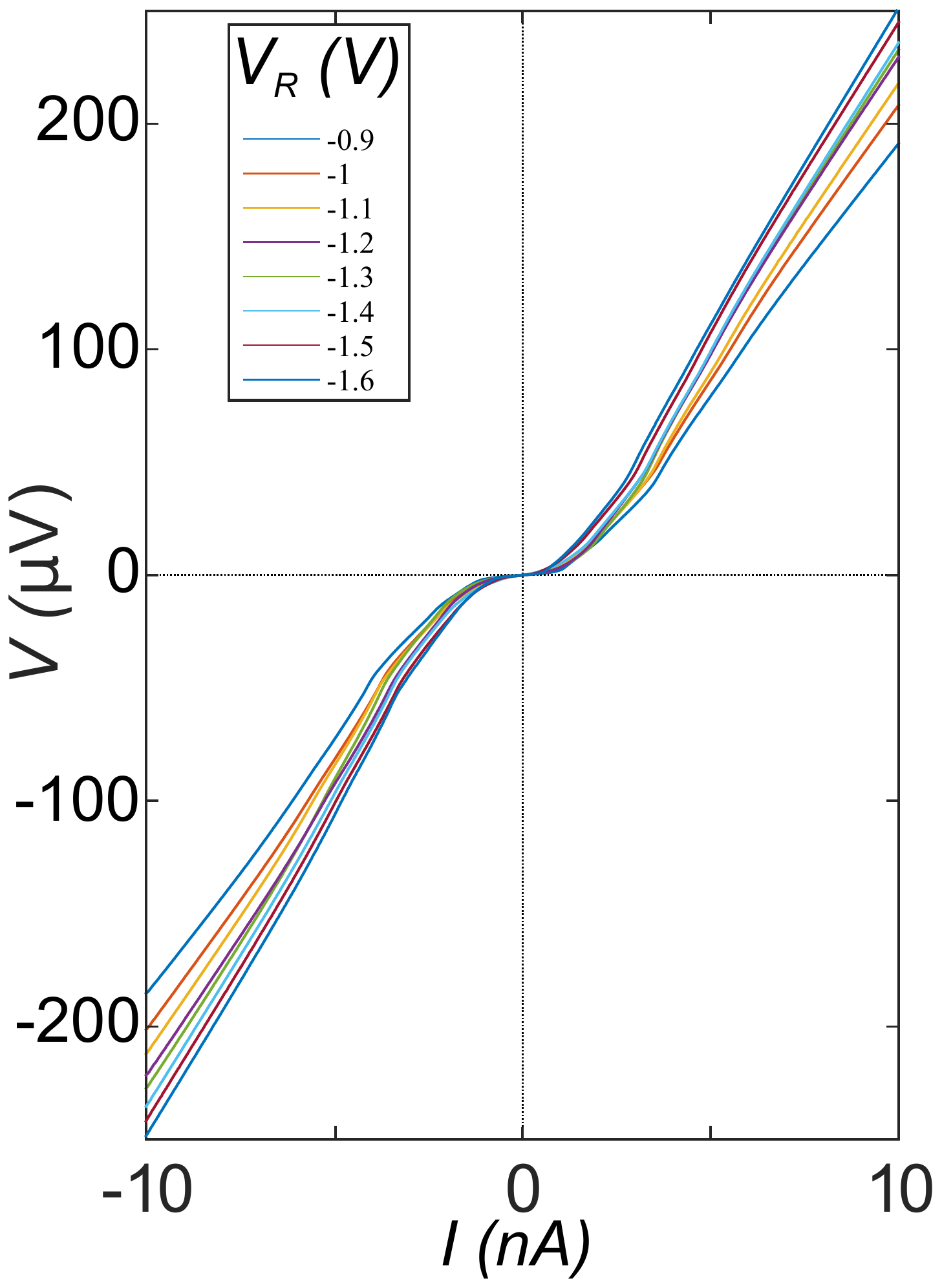}
	\caption{\textbf{Additional data} from the weak link regime (i) discussed in the main text, showing a set of full V-I curves for different V$_\text{R}$. 
	}
	\label{fig:WL}
\end{figure}

\newpage

\subsection{Calculation of Z, $\tau$ and I$_c$ in the weak link regime (i)}

We use the Blonder-Tinkham-Klapwjik (BTK) formalism described in Ref. \cite{blonder1982transition} to calculate the barrier parameter Z from the excess current and the high bias conductance.
The excess current $I_\text{exc}$ at an S-S interface is related to the Z parameter by

\begin{equation}
I_{exc} = 2\frac{g_n}{e(1-B(\infty))} \times \int_0^\infty dE (A(E)-B(E)+B(\infty)),
\end{equation}
with
\begin{eqnarray}
&A=\frac{\Delta^2}{E^2+(\Delta^2-E^2)(1+2Z^2)^2}, \\
&B=1-A,
\end{eqnarray}
for $E<\Delta$, and
\begin{eqnarray}
&A=\frac{u_0^2 v_0^2}{\gamma^2},\\
&B=\frac{(u_0^2-v_0^2)^2Z^2(1+Z^2)}{\gamma^2},
\end{eqnarray}
for $E>\Delta$. Furthermore, $B(\infty)=\frac{Z^2}{1+Z^2}$, and

\begin{eqnarray}
&u_0 = \frac{1}{2}(1+((E^2-\Delta^2)/E^2)^{1/2}, \\
&v_0 = 1-u_0^2,\\ 
&\gamma= (u_0^2 +Z^2(u_0^2-v_0^2))^2).
\end{eqnarray}

\begin{figure}(h)
	\includegraphics[width=0.4\linewidth]{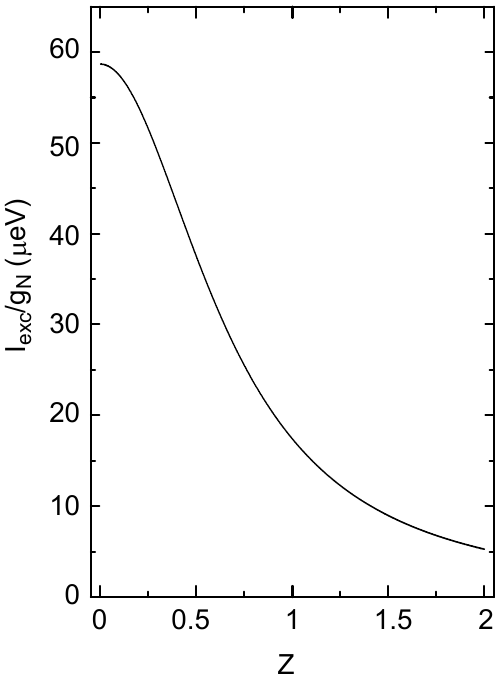}
	\caption{\textbf{Calculated Z-parameter: }Experimentally determined $I_\text{exc}/g_n$ and the corresponding Z-parameter determined for an S-S interface using BTK theory.}
	\label{fig:Z}
\end{figure}

We determine $I_\text{exc}$ from the experimental data by extrapolating the high bias conductance $g_n$ at I = 9 nA towards V=0. This yields the data shown in the top panel of Fig. 4(e) in the main text. Combining this with the respective $g_n$ for each gate voltage and using $\Delta$=22~$\mu$eV, as extracted from the dI/dV vs V curves, allows us to calculate the corresponding Z parameter, which leads to the curve shown in Fig. \ref{fig:Z}. By comparison we are then able to find the Z parameter as a function of $V_R$ and calculate the corresponding transmission coefficient in the normal state, $\tau = 1/(1+Z^2)$ \cite{blonder1982transition}.

For calculation of the critical current $I_c$ of a disordered point contact in the diffusive transport regime we apply the equation provided by Beenakker \cite{beenakker1992three}

\begin{equation}
I_c = 1.32\frac{\pi \Delta}{2e} \langle G\rangle
\end{equation}

where we have used the experimentally determined $g_n$ for the average conductance $\langle G\rangle$ and $\Delta=22~\mu$eV. This yields the blue dashed curve shown in the bottom panel in Fig.4(e) in the main text.

\newpage

\subsection{Estimating the size of the island}

\subsubsection{Numerical simulations of the electrostatic environment in the constriction}
We model the dielectric environment of the constriction by finite elements analysis. For that purpose we developed a 3D model of our device in COMSOL\textregistered 5.2. The simulation is developed similarly to that reported in \cite{monteiro2017side}, based on the electrostatic module.
The structure is modeled with 3 layers stacked on top of each other. At the top there is a 5\,nm-thick LaAlO$_3$ layer, having a dielectric constant of 24 \cite{krupka1994dielectric} and insulating character. The middle layer is the 2DES, modelled as a 10-nm thick metal with conductivity calculated from the experimental data. The bottom layer is the 1\,$\mu$m-thick SrTiO$_3$, which is an insulator with a field-dependent dielectric constant described by the Landau-Ginsburg-Devenshire Theory \cite{landau1984electrodynamics,ang2004dc}:

\begin{equation}
\epsilon_{STO} = 1+ \frac{B}{[1+(E/E_0)^2]^{2/3}},
\end{equation}

with E being the local electric field, $B = 25000$ and $E_0 = 82000$ V/m \cite{stornaiuolo2014weak}.
The split gates L and R are modelled as 100 nm thick triangular Au electrodes. The tips of the split gate are separated from each other by the distance $D=150$ nm. Gate voltages $V_L$ and $V_R$ are applied to the respective gates with respect to the drain reservoir, which is kept at ground potential. The source reservoir is voltage biased.
The island is modelled as a conductive disc with diameter $d$, which is separated from source and drain by gaps of width $g$. 

\begin{figure}[]
	\centering
	\includegraphics[width=0.6\linewidth]{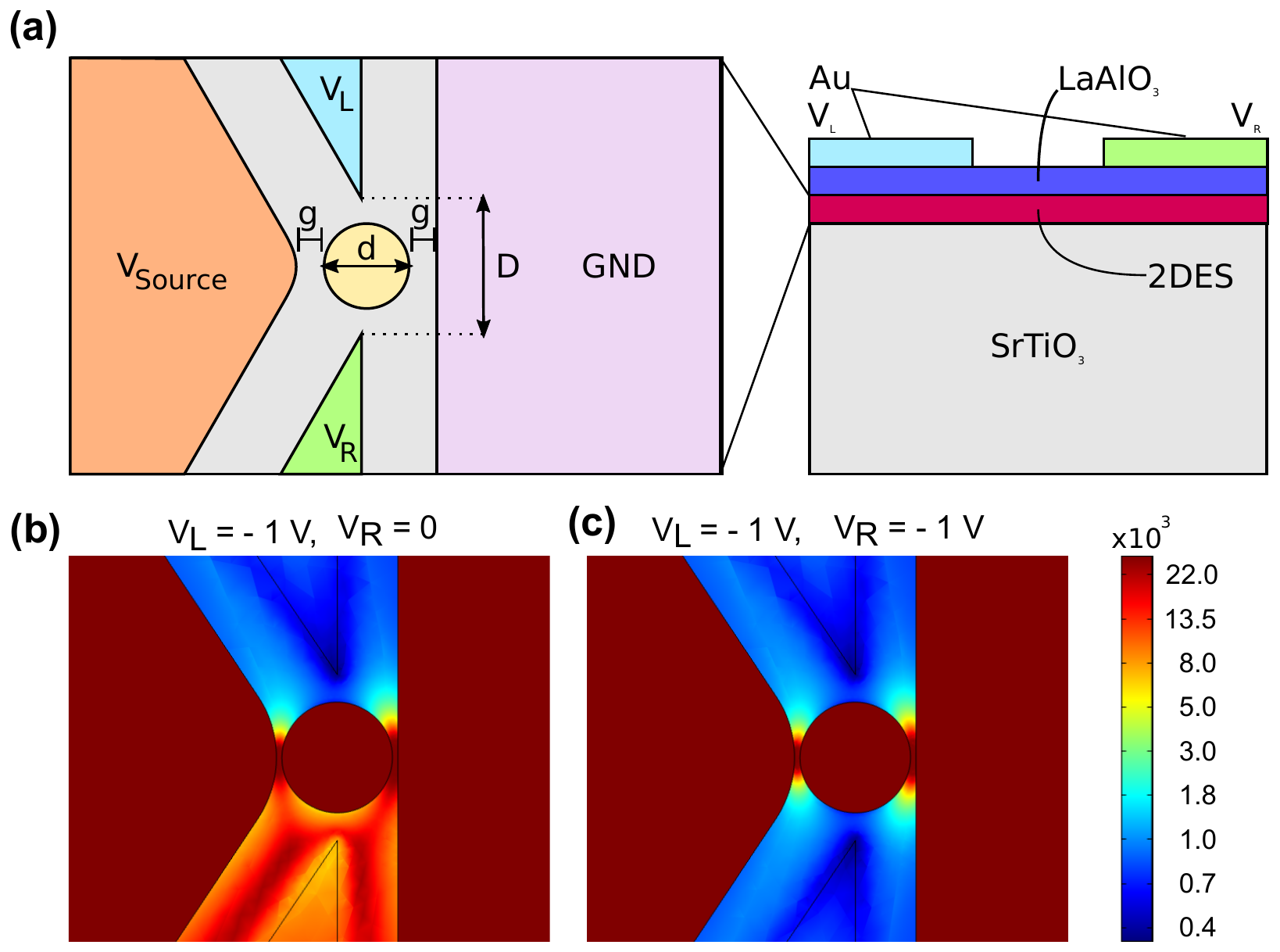}
	\caption{\textbf{Finite element analysis model of the device.} \textbf{(a)} Layer stack and geometry.\textbf{(b)} spacial map of the dielectric constant in the constriction at low temperature for gate voltages $V_\text{L}=-1$V, $V_\text{R}=0$V and \textbf{(c)} for $V_\text{L}=-1$V, $V_\text{R}=0$V.}
	\label{fig:Model}
\end{figure}

Figure \ref{fig:Model}(b) shows the spatial map of $\epsilon_r$ at the LAO/STO interface for $V_L = -1V$ and $V_R=0$. This visualizes the huge $\epsilon_r$ in large parts of the sample. In Fig.\ref{fig:Model}(c) the $\epsilon_r$ -map is shown for both gates at the same voltage ($V_L = V_R=-1V$). This results in reduction of the dielectric constant by more than one order of magnitude in the constriction and thus in the vicinity of the island.

\begin{figure}[]
	\centering
	\includegraphics[width=0.7\linewidth]{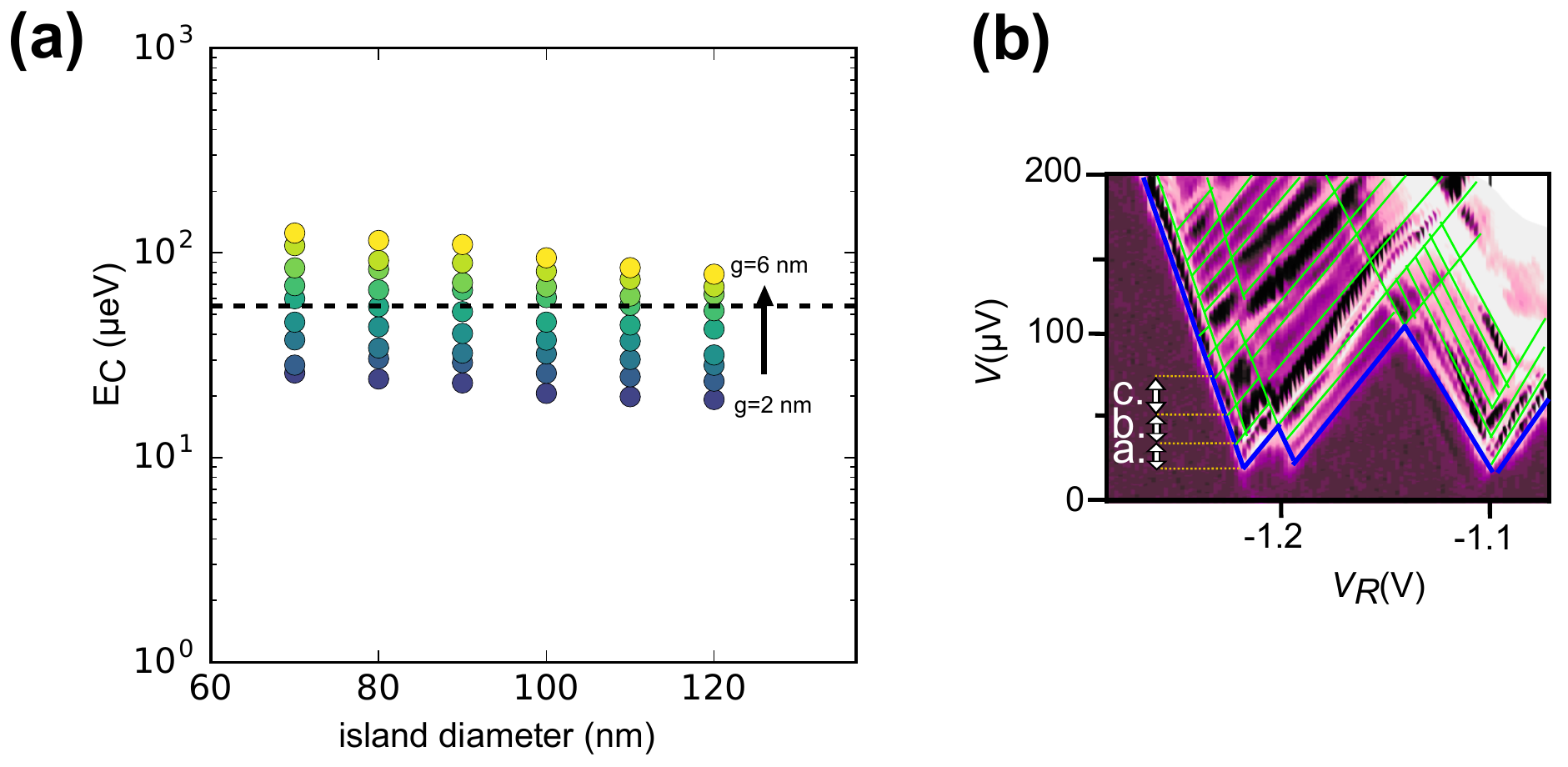}
	\caption{\textbf{Analysis of the Coulomb Diamonds:} \textbf{(a)} Results of the numerical simulation for $V_\text{L}=-1$V, $V_\text{R}=-1V$V showing the single electron charging energy as a function of island diameter. Colors indicate the assumed gap size ranging from $g=2$ to $6$ nm. \textbf{(b)} Close-up of the conductive region between two adjacent Coulomb diamonds, extracted from the main text. The borders of the diamonds are denoted with blue. Green lines indicate transport signatures of excited states on the island due to quantum confinement. The energy separation between different states are indicated with $a=14~\mu$V, $b= 17~ \mu $eV, $c = 21 ~\mu $eV.}
	\label{fig:Model2}
\end{figure}

The single electron charging energy of the island $E_C$ is extracted by calculating the voltage between source and drain that is required to change the polarization on the island by $e$ (electronic charge) while $V_{L,R} = -1 V$.
This calculation is carried out for different values of the island's diameter $d$ and the gap $g$. 
The result is shown in Fig.~\ref{fig:Model2}(a). Charging energies of 60 $\mu$eV [marked by the dashed line in Fig.~\ref{fig:Model2}(a)] to 100 $\mu$eV are obtained for island diameters of 70 nm to 120 nm if one allows $g$ to vary between approximately 3 nm and 6 nm.

\subsubsection{Analysis of excited states signatures}

We can further use the transport signatures of (excited) quantum states observed in the conductance diamond regime (iii) to estimate the size of the island.

A close up of the voltage range where these features are observed is depicted in Fig.~\ref{fig:Model2}(b). Only positive bias voltages are shown.
The delimiting lines of the Coulomb diamonds are indicated with blue lines.
Transport signatures of excited states of the island due to quantum confinement can be observed in the region between two adjacent Coulomb diamonds. They appear as lines of enhanced conductance which run in parallel with the borders of the Coulomb diamonds \cite{hanson2007spins,zwanenburg2013silicon}. In Fig.~\ref{fig:Model2}(b) they are denoted with green lines. The separation between these lines along the (vertical) bias voltage axis indicates their difference in energy, $\delta \varepsilon$. As an example, the energy separation of 4 such lines is indicated in Fig.~\ref{fig:Model2}(b) with $a=14~\mu$eV, $b=17~\mu$eV, $c=21~\mu$eV. 
Using a simple particle-in-a-box picture, we can estimate the spatial dimension $d$ required to obtain quantization energies of this order,

\begin{equation}
d=\sqrt{\frac{h^2}{8m\delta \varepsilon}}, 
\end{equation}

where $m = 0.7m_e$ is the effect electron mass in the LAO/STO 2DES, $m_e$ is the the bare electron mass and h Planck's constant.
Approximating by using $\delta \varepsilon = 20 \mu eV$ yields an island radius of approximately 80 nm. This is in the same range as the result obtained from the purely electrostatic considerations above.

This analysis clearly shows that the energy scale of the conductance diamonds is at least a factor 3 larger than that observed for the electronic orbital contributions originating from quantum confinement. Moreover, the numerical simulations clearly show that that Coulomb repulsion cannot be neglected in the device. The results strongly suggests that Coulomb blockade is the main contribution to the observed conductance diamonds. 
Using both signatures to estimate island size independently yields consistent results.

\newpage

\subsection{Coulomb diamond regime (iii) for large magnetic field}

\begin{figure}[]
	\centering
	\includegraphics[width=0.6\linewidth]{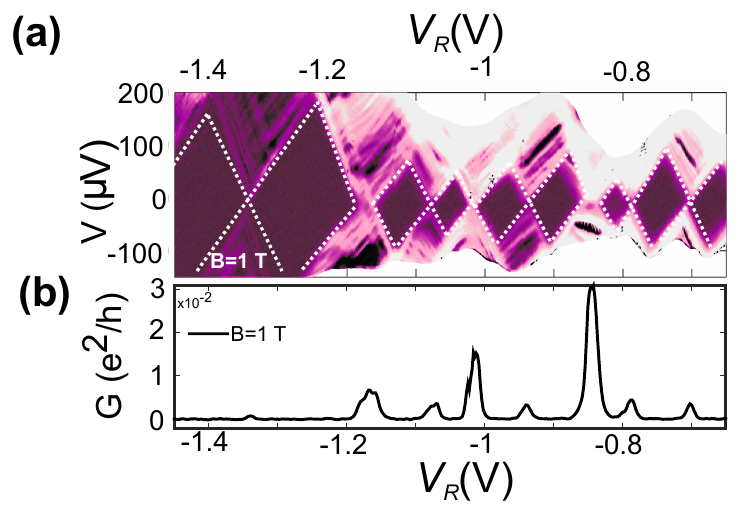}
	\caption{\textbf{Coulomb diamonds in a magnetic field: }\textbf{(a)} Coulomb diamonds discussed in the main text with a perpendicular magnetic field B=1T applied. It can be seen that the gap $V_\text{gap}$ has vanished. \textbf{(b)} Zero bias conductance as a function of V$_R$ extracted from (a).}
	\label{fig:Bfield}
\end{figure}

Figure \ref{fig:Bfield} (a) shows the series of Coulomb diamonds discussed in the main text with a perpendicular magnetic field B=1 T applied. Since $B$ is much larger than the typical critical magnetic field in superconducting LAO/STO 2DES, $B_c \approx 0.2T$, superconducting transport in the leads is suppressed. Therefore, the voltage gap $V_\text{gap} \approx 30~\mu$eV observed for $B=0$ vanishes.
The zero bias conductance is given in Fig.~\ref{fig:Bfield}(b). Interestingly, the amplitudes appear to alternate in an odd-even manner, 

resembling the parity effect in superconducting islands with two normal electrodes due to quasi-particle poising \cite{hergenrother1994charge} or photon assisted tunneling processes from radiation leaking though an imperfect shielding \cite{hergenrother1995photon}. This could suggest that despite the high magnetic field paired electrons are still present on the island. In the light of recent publications \cite{cheng2015electron,cheng2016tunable,tomczyk2016micrometer} this could hint at another signature of electron pairing without macroscopic superconductivity in LAO/STO.

\newpage

\subsection{Negative differential resistance and sub gap features in the Coulomb diamond regime (iii)}
\begin{figure}[]
	\centering
	\includegraphics[width=0.7\linewidth]{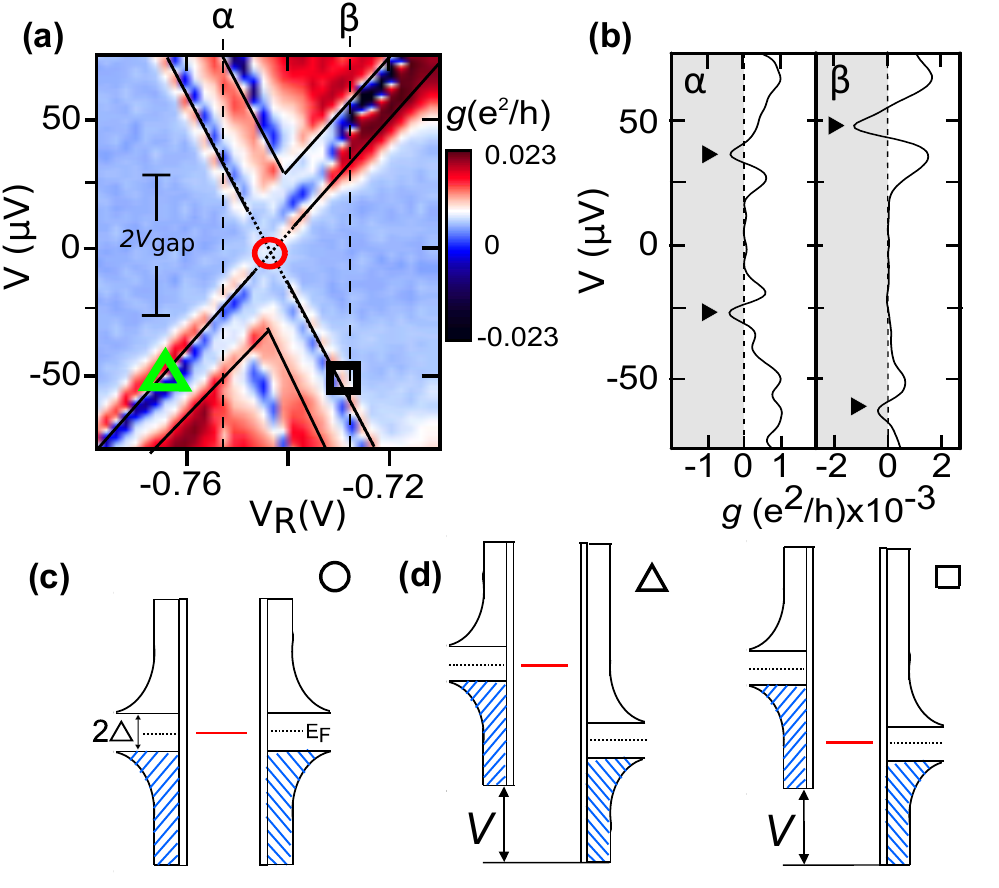}
	\caption{ \textbf{Negative differential resistance in the Coulomb diamonds of regime (iii)} \textbf{(a)} a close-up from Fig.4a) in the main text around $V_R=-0.75$ V. $\alpha$ and $\beta$ denote vertical line cuts shown in (b) which highlight the NDC features occurring symmetric with respect to bias voltage. The symbols ($\bigcirc$, $\Box$,$\triangle$) indicate the energy level configurations sketched in (c) and (d). 
	}
	\label{fig:X}
	
\end{figure}

\
We observe clear signatures of negative differential conductance (NDC) at the edges of the Coulomb diamonds. They occur for both positive and negative bias voltage, which indicates the presence of a superconducting gap in both reservoirs. As an example, a close up of the X shaped structure observed around $V_R=-0.75$ V is presented in Fig.\ref{fig:X}(a). Similar features have been observed recently by Cheng et al \cite{cheng2016tunable}, who pointed out a connection to tunable electron-electron interactions on LAO/STO quantum dots. It can clearly be seen from Fig.\ref{fig:X}(a) that the four 'arms' of the X intersect around V=0 ($\bigcirc$). 
This suggests that conductance originates from alignment of the island state with states close to the Fermi levels in the reservoirs (cf. the corresponding energy diagram in \ref{fig:X}(c). Each of the 'arms' exhibits a pronounced negative differential conductance (NDC). This becomes highlighted in Fig.\ref{fig:X}(b) where traces of $g$ are shown that are obtained from the vertical line cuts denoted $\alpha$ and $\beta$ in Fig.\ref{fig:X}(a). When going from V=0 towards positive or negative bias, $g$ exhibits first a positive peak at the intersections with the X structure, followed by a change of sign and a subsequent negative signal of approximately equal magnitude [black arrows in Fig.\ref{fig:X}(b)].

Generally, NDC reflects the alignment of the island state with a sharp DOS peak in the reservoirs. This becomes clear when we consider the energy diagrams presented in Fig.\ref{fig:X}(d) which sketch the configurations indicated with $\triangle$ and $\Box$ in Fig.~\ref{fig:X}(a). For both configurations increasing the bias voltage misaligns the island state with the DOS peak in the reservoir and thus reduces the current, leading to NDC. From the symmetric occurrence of NDC with respect to both $V$ and $V_R$ we infer that both reservoirs exhibit a sharp DOS peak. The location of the peaks close to the Fermi level could indicate that it is related to the Cooper pair DOS. Recent tunneling experiments, on the other hand, show indications of quasi particle sub gap states \cite{kuerten2017in}, which might give rise to features similar to the ones observed here.
We note that NDC at the boundaries of the Coulomb diamonds disappears if a magnetic field is applied, cf. Fig.\ref{fig:Bfield}. This further confirms that the NDC observed here originates from superconductivity.

\newpage

\subsection{Pre-Characterization of the Gates}

\begin{figure}[]
	\centering
	\includegraphics[width=0.8\linewidth]{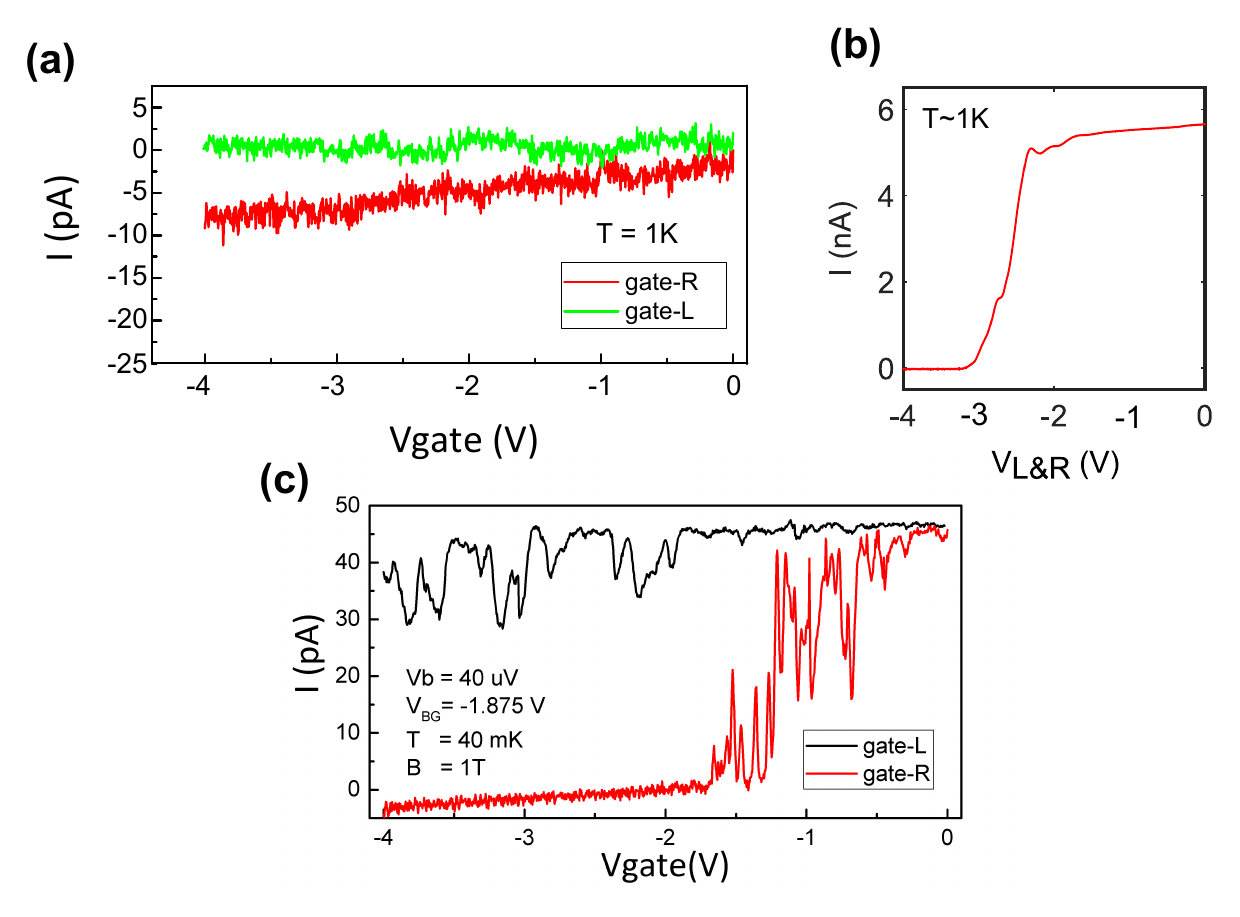}
	\caption{\textbf{Pre-characterization of the gates:} (a) Leakage currents of the gates measured at T=1K. (b) Pinch off curve at 1K for both gates L and R tuned simultaneously at 1K. \textbf{(c)} Pinch off curves when a voltage is applied only to one of the gates while the respective other one is grounded.}
	\label{fig:Gates}
\end{figure}

Figure~\ref{fig:Gates} (a) shows measurements of the gate leakage currents I$ _{leak}$ measured at $T\sim 1$ K by recording the dc drain current while varying the voltage applied to the respective gate. We find that even at V$_\text{L,R} = -4$ V the leakage current remains small, $|I_{leak}| < 10$ pA, even at -4 V.

In order to characterize the influence of the gate, we bias the 2DES with a constant voltage ($ V $=1 mV) and measure the drain current (I$ _{D} $) while changing the applied gate voltage. Note that this was done at finite back gate voltage to reduce the global carrier density such that the insulating state (pinch-off) is reached within the available gate voltage range. 
Figure \ref{fig:Gates} (b) shows the result obtained for applying a voltage to gates L and R simultaneously.
At low temperature, we observe the behaviour shown in Fig. ~\ref{fig:Gates}(c). Varying gate R alone enables us to block transport through the channel. Gate L, in contrast, does not pinch off the current, even at $V_{L}=-4V$. This indicates that the current does not flow through the 2DES underneath gate L, even at $V_L$=0, suggesting that it is depleted already at this stage and hence, the weak effect of gate L on transport.
Note that superconductivity is suppressed in these measurements due to a perpendicular magnetic field

\bibliography{Bibliography_full}

\end{document}